\documentclass[aps,prb,superscriptaddress,floatfix, twocolumn]{revtex4-1}

\usepackage{graphicx,graphics}
\usepackage{dcolumn}
\usepackage{amsmath,amssymb,amsfonts}
\usepackage{latexsym,verbatim}
\usepackage{bm}
\usepackage{color}
\usepackage[breaklinks=true,colorlinks,citecolor=blue,linkcolor=blue,urlcolor=blue]{hyperref}
\usepackage{ulem}

\def\be{\begin{equation}}
\def\ee{\end{equation}}
\def\ber{\begin{eqnarray}}
\def\eer{\end{eqnarray}}

\begin{document}
\title{The quasiparticle lifetime in a doped graphene sheet}

\author{Marco Polini}
\affiliation{NEST, Istituto Nanoscienze-CNR and Scuola Normale Superiore, I-56126 Pisa, Italy}
\author{Giovanni Vignale}
\affiliation{Department of Physics and Astronomy, University of Missouri, Columbia, Missouri 65211, USA}

\begin{abstract}
We present a  calculation of the quasiparticle decay rate due to electron-electron interactions in a doped graphene sheet. In particular, we emphasize subtle differences between the perturbative calculation of this quantity in a doped graphene sheet and the corresponding one in ordinary parabolic-band two-dimensional (2D) electron liquids. In the random phase approximation, dynamical overscreening near the light cone yields a universal quasiparticle lifetime, 
which is independent of the dielectric environment surrounding the 2D massless Dirac fermion fluid.
\end{abstract}

\maketitle

\section{Introduction}

Gabriele Giuliani loved the Landau theory of normal Fermi liquids~\cite{Nozieres,Pines_and_Nozieres,shankar_rmp_1994,Giuliani_and_Vignale}.  The notion that a system of strongly interacting particles could behave like an ideal gas of plain non-interacting particles, was to him a source of endless fascination.  This was largely a reflection of his  ``down-to-earth" approach to theoretical physics.  Gabriele disliked all forms of mystification and particularly the widespread one of couching trivial or wrong ideas in high-sounding theoretical language.  Fermi liquid theory, with its deceptive simplicity, was precisely the opposite of mystification: it was the sophisticated plainness he was striving for.
At the heart of Fermi liquid theory lies the concept of  ``quasiparticle" -- a quasi-exact  eigenstate of a single excited particle that decays very slowly in time.  How slowly?  The critical requirement is that the decay rate of the state  remain much smaller than its  energy  in the limit that the latter tends to zero.  If this condition is satisfied, then an ``adiabatic switching-on" process  becomes viable, whereby, starting from an infinitely long-lived excited eigenstate of the non-interacting system, and slowly turning on the interaction  (``slowly" meaning at a rate that is much longer than the excitation frequency -- yet faster than the decay rate), one generates the long-lived eigenstate of the interacting system.

A standard argument for estimating the decay rate (also known as inverse lifetime) of a quasiparticle goes as follows.  {\it Assuming that long-lived quasiparticles exist} with a small energy $\xi$ in the vicinity of the Fermi surface it is evident that they can only decay by scattering into other available (i.e., empty) quasiparticle states.  This is because Pauli's exclusion principle pre-empts scattering of a fermion into an occupied state (we ignore spin for simplicity).  The number of  available states is thus proportional to $\xi$ (at zero temperature) or to $T$, if $\xi\ll k_{\rm B}T$.  Further, conservation of momentum and energy require that the decay be accompanied by the production of a 
quasi-electron-quasi-hole pair, whose energy is also of the order of $\xi$ or $k_{\rm B}T$, whichever is larger.  The density of such pairs is proportional to $\xi$ or $k_{\rm B}T$.  Taking the two factors together, we conclude that the quasiparticle decay rate is proportional to $\xi^2$ or $(k_{\rm B} T)^2$, which is indeed much smaller than the excitation frequency, $\xi$ or $k_{\rm B}T$, in the limit that the latter tends to zero. 

Notice that this somewhat circular argument is valid (when it is valid) regardless of the strength of the electron-electron interaction.  And indeed, for three-dimensional Fermi systems  the naive argument gives the right answer, even when the interactions are very strong (as in $^3$He and in heavy fermion compounds) and the renormalizations of the effective (i.e., quasiparticle) mass are correspondingly large.  The situation is completely different in one spatial dimension, where the same argument fails to predict the collectivization of the electron and the formation of the Luttinger liquid state (the situation is very well described in Giamarchi's book~\cite{Giamarchi}). 

What about the two-dimensional (2D) electron liquid?  In the early 1980s, when Gabriele was just beginning  his career, two-dimensional electron gases (2DEGs) in GaAs-based heterostructures and Si inversion layers were among the most fashionable systems  studied by condensed matter physicists.  The twin discoveries of the localizing effect of impurities in two spatial dimensions~\cite{abrahams_prl_1979} (scaling theory of localization) and, more subtly, of the quantum Hall effect~\cite{vonKlitzing_prl_1980}, which critically depended on the former,  appeared to undermine the Fermi liquid picture of the 2DEG.  The very existence of the metallic state of the 2DEG was in doubt~\cite{abrahams_prl_1979}.
With his ``no-nonsense'' attitude Gabriele followed those developments closely, but never bought into the most adventurous ideas.  To those who denied the existence of the metallic state of electrons in 2D GaAs he was likely to suggest the following thought experiment: ``{\it OK, let us stick this end of the sample into the power socket, while you hold the other end...}"  
But at the same time he would not accept uncritically the conventional wisdom about the Fermi liquid state in two spatial dimensions.  And it was so that, during his postdoc with John Quinn at Brown University, he began to investigate the key question of the quasiparticle lifetime in the 2DEG.  Working within the Fermi liquid picture, he was able to establish~\cite{giuliani_prb_1982} that the decay rate of a quasiparticle in the 2DEG does not scale as $\xi^2$ or $(k_{\rm B}T)^2$ as the naive argument would suggest, but rather as $-\xi^2\ln(\xi)$ or   $-(k_{\rm B}T)^2 \ln(k_{\rm B} T)$, depending on whether $k_{\rm B}T \ll \xi$ or $k_{\rm B} T\gg\xi$, respectively~\cite{otherearlierliterature}:
\be\label{eq:GiulianiQuinn}
\frac{1}{\tau_{\bm k}} =
\left\{
\begin{array}{l}
 {\displaystyle - \frac{\varepsilon_{\rm F}}{\hbar} \frac{1}{4\pi} \left(\frac{\xi_{\bm k}}{\varepsilon_{\rm F}}\right)^2\ln{\left(\frac{|\xi_{\bm k}|}{\varepsilon_{\rm F}}\right)},~{\rm for}~k_{\rm B}T \ll |\xi_{\bm k}|}\vspace{0.2 cm}\\
  {\displaystyle - \frac{\varepsilon_{\rm F}}{\hbar} \frac{1}{2\pi} \left(\frac{k_{\rm B}T}{\varepsilon_{\rm F}}\right)^2\ln{\left(\frac{k_{\rm B}T}{\varepsilon_{\rm F}}\right)},~{\rm for}~k_{\rm B}T \gg |\xi_{\bm k}|}
\end{array}
\right.~.
\ee
Here $\xi_{\bm k} = \hbar^2 {\bm k}^2/(2 m) - \varepsilon_{\rm F}$ is the parabolic-band energy measured from the Fermi energy $\varepsilon_{\rm F}$, $m$ being the electron's (band) mass and $\hbar{\bm k}$ the 2D momentum. 
The unexpected logarithmic enhancement of the decay rate is due to a subtle feature of the 2D phase space available for the scattering of  quasi-particles near the Fermi surface---a feature that is not captured by the naive argument. Another surprising feature of the Giuliani-Quinn formula for the decay rate
is that the coefficient of the leading terms $-\xi^2 \ln(\xi)$ or $-T^2\ln(T)$ is {\it independent} of the electron-electron coupling constant or, as Giuliani and Quinn aptly put it, of the magnitude of the electron charge. 
This counterintuitive feature arises from the fact that, in the Giuliani-Quinn theory, the dominant contribution to the decay rate arises from scattering processes with small momentum transfer $q$:  these are the processes for which the Coulomb interaction between two quasiparticles is most strongly screened~\cite{Giuliani_and_Vignale} by the electronic medium that surrounds them, leading to an effective interaction that depends only on the non-interacting density of states. 

Eq.~(\ref{eq:GiulianiQuinn}) provides the justification for applying Fermi liquid theory to the 2DEG, at least when disorder is not too strong. 
The logarithmic enhancement of the decay rate does not create any  serious danger to the stability of  quasiparticles,  probably less than  Gabriele's thought experiment to its hypothetical subjects.
Over the years, the paper~\cite{giuliani_prb_1982} in which Eq.~(\ref{eq:GiulianiQuinn}) was first reported grew to be the  standard reference on the subject. 
Adjustments had to be made~\cite{adjustments} over the years to include the contributions of $2k_{\rm F}$ scattering, vertex corrections, exchange effects, etc..., but none of these refinements changed the basic picture established in the original paper.  Furthermore, numerous experiments since then have established the validity of the Fermi liquid concept in the 2DEG~\cite{fermiliquidrecentexperiments}, the quasiparticle lifetime has been probed in detail~\cite{murphy_prb_1995},  and the existence of the metallic state has been demonstrated~\cite{kravchenko_prb_1994}.

Fast-forward 40$^+$ years to 2004,  the year in which, for the first time, few-layer graphene sheets were electrically contacted and the field effect was demonstrated~\cite{novoselov_science_2004}.
In its pristine state, graphene, i.e.~a single layer of Carbon atoms arranged in a honeycomb structure, is a semimetal~\cite{graphene}. Its conical conduction and valence bands have dispersions $\sim \pm \hbar v_{\rm F}(k-k_{\rm D})$ in the vicinity of the Dirac point $k_{\rm D}$, where they touch.  Due to the vanishing density of states at the Fermi level one would expect a complete break-down of the Fermi liquid paradigm.  And, indeed, many-body calculations~\cite{gonzalezetal} suggest that the so-called massless Dirac fermion (MDF) quasiparticles exhibit singular features, such as a logarithmically diverging velocity~\cite{elias_naturephys_2011} and linear-in-energy decay rates~\cite{siegel_pnas_2011}, which are hardly compatible with the Landau Fermi liquid paradigm.
Nevertheless, these singularities are found to be relevant only for extremely low carrier densities and, when a sizeable Fermi surface is created (by doping, or, more conveniently, by electrostatic gating),  the  conventional Fermi liquid description seems to take hold again, even in suspended sheets, where the strength of electron-electron interactions is the largest. To be convinced that this is truly the case, one 
must  calculate carefully the decay rate for quasiparticles near the Fermi surface.  One might suppose that the presence of the Fermi surface erases any difference between the ordinary Schr\"{o}dinger electrons of a 2DEG 
and the MDFs of graphene: after all the parabolic dispersion of Schr\"{o}dinger electrons is approximately linear in the vicinity of the Fermi surface.  However,  the lesson of the Giuliani-Quinn paper is that such a-priori arguments must be taken with a good dose of skepticism, because subtle differences in the structure of the phase space can lead to quantitative differences in the decay rate.
And indeed, a careful calculation, presented in the next few sections, exposes several differences between the calculation of the quasiparticle lifetime in graphene and in the 2DEG---differences that arise from the suppression of backscattering (characteristic of MDFs) as well as from the large enhancement of screening in MDF systems at frequencies near the light cone $\omega = \pm v_{\rm F} q$. The final upshot of the calculation, however,  is that the 
Giuliani-Quinn picture remains valid, with the added feature that collinear scattering processes with small momentum transfer are now more important than ever, and completely dominate the behavior of the quasiparticle lifetime, while $2k_{\rm F}$ processes (initially neglected by Giuliani and Quinn) are happily suppressed.

The Fermi liquid properties and Coulomb decay rates of quasiparticles in graphene sheets have been studied by many authors. We have provided a (certainly incomplete) list of pertinent works in Refs.~\onlinecite{barlas_prl_2007,polini_ssc_2007,hwang_prb_2007,dassarma_prb_2007,polini_prb_2008,hwang_prb_2008,schutt_prb_2011,tomadin_prb_2013,li_prb_2013,song_prb_2013,basko_prb_2013,briskot_prb_2014,brida_naturecommun_2013,tielrooij_naturephys_2013}. 
In this Article we present a pedagogical description of the calculation leading to an explicit formula for the Coulomb decay rate (i.e.~inverse lifetime) of a weakly-excited plane-wave state in a doped graphene sheet. 
Our main results, Eq.~(\ref{eq:PW_lifetime_zeroT}) and Eq.~(\ref{eq:PW_lifetime_finiteT}), have been derived earlier by other authors (see, e.g., Ref.~\onlinecite{li_prb_2013}): the emphasis of this work is 
on the intermediate steps of the calculation.

\section{Coulomb-enabled two-body decay rates in a doped graphene sheet}

In this Section we present a theory of the decay rate $1 / \tau_{{\bm k}, \lambda}$ 
of a plane-wave state with momentum $\hbar {\bm k}$ and band index $\lambda$ in a doped graphene sheet, 
at a temperature $T$. We will consider decay rates solely due to two-body Coulomb collisions.

For future purposes, we introduce the so-called graphene's fine-structure constant~\cite{kotov_rmp_2012} $\alpha_{\rm ee}$,
\be
\alpha_{\rm ee} = \frac{e^2}{\epsilon \hbar v_{\rm F}}~.
\ee
Here, the dielectric constant $\epsilon$ is the average of the dielectric constants $\epsilon_1$ and $\epsilon_2$ 
of the media above and below the graphene flake, i.e.~$\epsilon \equiv (\epsilon_1 + \epsilon_2)/2$. The dimensionless parameter $\alpha_{\rm ee}$ determines the strength of electron-electron interactions with respect to the kinetic energy. 

We start by considering the so-called ``G$_0$W-RPA'' approximation for the imaginary part of the self-energy $\Sigma_{\lambda}({\bm k}, \omega)$ in a doped graphene sheet~\cite{polini_ssc_2007} (from now on we set $\hbar =1$, unless otherwise stated):
\begin{widetext}
\be\label{eq:selfenergyg0w}
\Im m[\Sigma_{\lambda}({\bm k}, \omega)] =- \int \frac{d^2{\bm q}}{(2\pi)^2}\sum_{\lambda'}\Im m\left[\frac{v_q}{\varepsilon(q, \omega - \xi_{{\bm k} - {\bm q},\lambda'}, T)}\right]~{\cal F}_{\lambda\lambda'}(\theta_{{\bm k},{\bm k} - {\bm q}}) [ n_{\rm B}(\omega - \xi_{{\bm k} - {\bm q},\lambda'}) + n_{\rm F}(-\xi_{{\bm k} - {\bm q},\lambda'})]~,
\ee
\end{widetext}
where $\lambda, \lambda'$ are band indices ($\lambda=+$ denotes conduction-band states, $\lambda= -$ denotes valence-band states), $\theta_{{\bm k},{\bm k} - {\bm q}}$ is the angle between ${\bm k}$ and ${\bm k} - {\bm q}$,
\begin{equation}
{\cal F}_{\lambda \lambda'}(\varphi) \equiv \frac{1 + \lambda\lambda'\cos{(\varphi)}}{2}
\end{equation}
is the usual chirality factor, and
\begin{equation}
\xi_{{\bm k}, \lambda} \equiv \varepsilon_{{\bm k}, \lambda} - \mu =  \lambda v_{\rm F} k -\mu
\end{equation}
are Dirac-band single-particle energies measured from the chemical potential $\mu$.

In Eq.~(\ref{eq:selfenergyg0w}) $n_{{\rm B}/{\rm F}}(x) \equiv 1/[\exp(\beta x) \mp 1]$ are the usual Bose (Fermi) statistical factors with $\beta = (k_{\rm B} T)^{-1}$ and
\be
\varepsilon(q, \omega, T) \equiv 1 - v_{q} \chi^{(0)}(q, \omega, T)
\ee
is the finite-temperature dynamical screening function in the random phase approximation (RPA)~\cite{Giuliani_and_Vignale}. Here $v_q = 2\pi e^2/(\epsilon q)$ is the 2D Fourier transform of the Coulomb interaction and $\chi^{(0)}(q, \omega, T)$ the non-interacting finite-temperature density-density response function of a 2D gas of MDFs~\cite{ramezanali_jphysa_2009}. It contains both intra- and inter-band contributions.

Note that
\be
\Im m\left[\frac{1}{\varepsilon(q, \omega, T)}\right] = v_{q}  \frac{\Im m[\chi^{(0)}(q, \omega, T)] }{ |\varepsilon(q, \omega, T)|^{2}}~.
\ee
Using the previous identity in Eq.~(\ref{eq:selfenergyg0w}) we find the following expression for the decay rate due to two-body Coulomb collisions:
\begin{widetext}
\ber\label{eq:lifetime}
\frac{1}{\tau_{{\bm k}, \lambda}}&\equiv& 2\Im m[\Sigma_{\lambda}({\bm k}, \xi_{{\bm k}, \lambda})]= - 2\sum_{\lambda'}\int \frac{d^2{\bm q}}{(2\pi)^2}v^2_q~\frac{\Im m[\chi^{(0)}(q,\xi_{{\bm k}, \lambda} - \xi_{{\bm k} - {\bm q},\lambda'},T)]}{|\varepsilon(q, \xi_{{\bm k}, \lambda} - \xi_{{\bm k} - {\bm q},\lambda'}, T)|^2}~{\cal F}_{\lambda\lambda'}(\theta_{{\bm k},{\bm k} - {\bm q}}) \nonumber\\
&\times& [ n_{\rm B}(\xi_{{\bm k}, \lambda} - \xi_{{\bm k} - {\bm q},\lambda'}) + n_{\rm F}(-\xi_{{\bm k} - {\bm q},\lambda'})]~.
\eer
We now use the exact identity 
\be
n_{\rm B}(\xi_{{\bm k}, \lambda} - \xi_{{\bm k} - {\bm q},\lambda'}) + n_{\rm F}(-\xi_{{\bm k} - {\bm q},\lambda'}) = \frac{1- n_{\rm F}(\xi_{{\bm k} - {\bm q},\lambda'})}{1 - \exp[-\beta(\xi_{{\bm k}, \lambda} - \xi_{{\bm k} - {\bm q},\lambda'})]}
-\frac{n_{\rm F}(\xi_{{\bm k} - {\bm q},\lambda'})}{1 - \exp[\beta(\xi_{{\bm k}, \lambda} - \xi_{{\bm k} - {\bm q},\lambda'})]}
\ee
\end{widetext}
and introduce the following auxiliary delta function on the energy transfer $\omega$:
\be
1 = \int_{-\infty}^{\infty} d\omega~\delta(\xi_{{\bm k}, \lambda} - \xi_{{\bm k}-{\bm q}, \lambda'} - \omega)~.
\ee
We can therefore rewrite Eq.~(\ref{eq:lifetime}) as follows
\begin{widetext}
\ber\label{eq:quasielectron-quasihole-lifetime}
\frac{1}{\tau_{{\bm k}, \lambda}}&=&-\frac{2}{(2\pi)^2}\sum_{\lambda'}\int_{-\infty}^{+\infty}d\omega~\frac{1- n_{\rm F}(\xi_{{\bm k},\lambda} - \omega)}{1 - \exp(-\beta\omega)}\int_0^{+\infty}dq~q\left|\frac{v_q}{\varepsilon(q, \omega, T)}\right|^2\Im m[\chi^{(0)}(q,\omega,T)]A_{\lambda\lambda'}(k,q,\omega)\nonumber \\
&+&\frac{2}{(2\pi)^2}\sum_{\lambda'}\int_{-\infty}^{+\infty}d\omega~\frac{n_{\rm F}(\xi_{{\bm k},\lambda} -\omega)}{1 - \exp(\beta\omega)}\int_0^{+\infty}dq~q\left|\frac{v_q}{\varepsilon(q, \omega, T)}\right|^2\Im m[\chi^{(0)}(q,\omega,T)]A_{\lambda\lambda'}(k,q,\omega)~.
\eer
\end{widetext}
Note that the second term in the previous equation can be obtained from the first term by performing the replacements $1- n_{\rm F}(\xi_{{\bm k},\lambda} - \omega) \to n_{\rm F}(\xi_{{\bm k},\lambda} -\omega)$, $1 - \exp(-\beta\omega) \to 1 - \exp(\beta\omega)$, and changing the overall sign. For this reason, it is customary~\cite{Giuliani_and_Vignale} to define the first  term in Eq.~(\ref{eq:quasielectron-quasihole-lifetime}) as the quasiparticle decay rate, the second term as the quasihole decay rate and the sum of the two as the decay rate of the plane-wave state ${\bm k}, \lambda$:
\be
\frac{1}{\tau_{{\bm k}, \lambda}} \equiv \frac{1}{\tau^{({\rm e})}_{{\bm k}, \lambda}} + \frac{1}{\tau^{({\rm h})}_{{\bm k}, \lambda}}~.
 \ee
In Eq.~(\ref{eq:quasielectron-quasihole-lifetime}) we have  introduced the following angular integral
\ber\label{eq:angular_integral}
A_{\lambda\lambda'}(k,q,\omega) &\equiv& \int_0^{2\pi} d\theta~\delta(\xi_{{\bm k}, \lambda} - \xi_{{\bm k}-{\bm q}, \lambda'} - \omega) \nonumber \\
&\times&{\cal F}_{\lambda\lambda'}(\theta_{{\bm k},{\bm k} - {\bm q}})~,
\eer
where $\theta$ is the angle between ${\bm q}$ and ${\bm k}$, which can be oriented along the ${\hat {\bm x}}$ axis without loss of generality, i.e.~${\bm k}  = k {\hat {\bm x}}$. For future purposes it is important to note that
\be
\cos(\theta_{{\bm k}, {\bm k} - {\bm q}}) = \frac{k - q \cos(\theta)}{\sqrt{k^2 + q^2 - 2 kq \cos(\theta)}}~.
\ee

Since the integrand in Eq.~(\ref{eq:angular_integral}) is a function of $\cos(\theta)$ only, we can write 
\ber\label{eq:angular_integral_symmetry}
A_{\lambda\lambda'}(k,q,\omega) &=& 2\int_0^{\pi} d\theta~\delta(\xi_{{\bm k}, \lambda} - \xi_{{\bm k}-{\bm q}, \lambda'} - \omega) \nonumber \\
&\times&{\cal F}_{\lambda\lambda'}(\theta_{{\bm k},{\bm k} - {\bm q}})~.
\eer

The function $A_{\lambda\lambda'}(k,q,\omega)$ can be easily evaluated analytically. One first realizes that the delta function in Eq.~(\ref{eq:angular_integral}) gives a non-zero contribution to $A_{\lambda\lambda'}$ if and only if the equality
\be\label{eq:deltafunction}
v_{\rm F} \lambda k -  v_{\rm F} \lambda' \sqrt{k^2 + q^2 - 2 kq \cos(\theta)} =  \omega
\ee
is satisfied. This condition does not depend on the chemical potential $\mu$.

\subsection{Intra-band contribution}
\label{sect:intra-band}

For $\lambda' = +1$ (intra-band scattering) Eq.~(\ref{eq:deltafunction}) reduces to
\be\label{eq:firstconstraint}
\sqrt{k^2 + q^2 - 2 kq \cos(\theta)} = k - \frac{\omega}{v_{\rm F}}~,
\ee
which requires $k \geq \omega/v_{\rm F}$. When this condition is satisfied,
\be\label{eq:costhetaintraband}
\cos(\theta) = \frac{q^2 - \omega^2/v^2_{\rm F} + 2 k \omega/v_{\rm F}}{2 kq}~,
\ee
which in turn requires
\be\label{eq:secondconstraint}
\left|\frac{q^2 - \omega^2/v^2_{\rm F} + 2 k \omega/v_{\rm F}}{2 kq}\right| \leq 1.
\ee
Eq.~(\ref{eq:costhetaintraband}) admits always one solution in the interval $[0,\pi]$. When Eq.~(\ref{eq:costhetaintraband}) is satisfied,
\be
\cos(\theta_{{\bm k}, {\bm k} - {\bm q}}) \to 1 - \frac{q^2 - \omega^2/v^2_{\rm F}}{2k(k- \omega/v_{\rm F})}~, 
\ee
and therefore
\ber
{\cal F}_{++}(\cos(\theta_{{\bm k}, {\bm k} - {\bm q}})) &\to& 
1- \frac{q^2 - \omega^2/v^2_{\rm F}}{4k(k-\omega/v_{\rm F})} \nonumber\\
&\equiv& {\cal F}^\star_{++}(k,q,\omega)~.
\eer
Note that ${\cal F}^\star_{++}(k,q,\omega)=1$ for $\omega = \pm v_{\rm F}q$.

For intra-band scattering the result of the angular integration in Eq.~(\ref{eq:angular_integral_symmetry}) is therefore
\begin{widetext}
\ber\label{eq:angular_integral_MDF_intraband}
A_{++}(k,q,\omega) &=& 2\times \frac{2c (k - \omega/v_{\rm F}){\cal F}^\star_{++}(k,q,\omega)}{v_{\rm F}\sqrt{(
2 k + q - \omega/v_{\rm F})(2 k - q - \omega/v_{\rm F})(q - \omega/v_{\rm F})(q + \omega/v_{\rm F})}}\nonumber \\
&\times&\Theta(k - \omega/v_{\rm F})\Theta\left(1- \left|\frac{q^2 - \omega^2/v^2_{\rm F} +2 k (\omega/v_{\rm F})}{2kq}\right|\right)~,
\eer
\end{widetext}
where the first factor of two is the same as the one appearing in Eq.~(\ref{eq:angular_integral_symmetry}). In Eq.~(\ref{eq:angular_integral_MDF_intraband}) $\Theta(x) = 1$ if $x \geq 0$ and $0$ otherwise. 
Furthermore, $c$ is a numerical coefficient:
\be
c = 
\left\{
\begin{array}{l}
1/2,~{\rm for}~q = \omega/v_{\rm F}~{\rm and}~q = 2k - \omega/v_{\rm F}\vspace{0.2 cm}\\
1,~{\rm elsewhere}
\end{array}
\right.~.
\ee
Indeed, for $q = \omega/v_{\rm F}$ and $q = 2k - \omega/v_{\rm F}$ we have $\cos(\theta) = +1$---see Eq.~(\ref{eq:costhetaintraband})---and therefore the solution, i.e.~$\theta = 0$, falls on the boundary of the integration domain in Eq.~(\ref{eq:angular_integral_symmetry}) (and therefore the integral of the delta function gives an extra factor $1/2$).

A careful analysis of Eq.~(\ref{eq:angular_integral_MDF_intraband}) allows us to conclude that
\be\label{eq:final_angular_integral_MDF_intraband}
A_{++} =\frac{4c(k - \omega/v_{\rm F}){\cal F}^\star_{++}(k,q,\omega)}{v_{\rm F}\sqrt{[(2 k - \omega/v_{\rm F})^2 - q^2](q^2 - \omega^2/v^2_{\rm F})}}
\ee
for
\be\label{eq:final-inequality-intraband}
\frac{\omega}{v_{\rm F}} \leq k~~{\rm and}~~\frac{|\omega|}{v_{\rm F}} \leq q \leq 2k - \frac{\omega}{v_{\rm F}}~,
\ee
and zero elsewhere. Note that for $|\omega|/v_{\rm F} \leq q \leq 2k - \omega/v_{\rm F}$ the argument of the square root in Eq.~(\ref{eq:final_angular_integral_MDF_intraband}) is positive.
\subsection{Inter-band contribution}
\label{sect:inter-band}

For $\lambda' = -1$ (inter-band scattering) Eq.~(\ref{eq:deltafunction}) requires $0  \leq k \leq \omega/v_{\rm F}$ since it must be
\be\label{eq:firstconstraint_interband}
\sqrt{k^2 + q^2 - 2 kq \cos(\theta)} =\frac{\omega}{v_{\rm F}} - k~.
\ee
Following identical steps to those described in Sect.~\ref{sect:intra-band} we find
\begin{widetext}
\ber\label{eq:angular_integral_MDF_interband}
A_{+-}(k,q,\omega) &=& \frac{4c(\omega/v_{\rm F} - k){\cal F}^\star_{+-}(k,q,\omega)}{v_{\rm F}\sqrt{(
2 k + q - \omega/v_{\rm F})(2 k - q - \omega/v_{\rm F})(q - \omega/v_{\rm F})(q + \omega/v_{\rm F})}}\nonumber \\
&\times&\Theta(\omega/v_{\rm F} -k)\Theta\left(1- \left|\frac{q^2 - \omega^2/v^2_{\rm F} +2 k (\omega/v_{\rm F})}{2kq}\right|\right)~.
\eer
\end{widetext}
In Eq.~(\ref{eq:angular_integral_MDF_interband}) we have defined
\ber
{\cal F}^\star_{+-}(k,q,\omega)&=& \frac{q^2 - \omega^2/v^2_{\rm F}}{4k(k-\omega/v_{\rm F})}~.
\eer
Note that ${\cal F}^\star_{+-}(k,q,\omega)=0$ for $\omega = \pm v_{\rm F}q$. A careful analysis of Eq.~(\ref{eq:angular_integral_MDF_interband}) allows us to conclude that
\be\label{eq:final_angular_integral_MDF_interband}
A_{+-} = \frac{4c (\omega/v_{\rm F} - k){\cal F}^\star_{+-}(k,q,\omega)}{v_{\rm F}
\sqrt{[(2 k - \omega/v_{\rm F})^2 - q^2](q^2 - \omega^2/v^2_{\rm F})}}~.
\ee
for
\be
\frac{\omega}{v_{\rm F}} \geq k~~{\rm and}~~\left|2 k - \frac{\omega}{v_{\rm F}}\right| \leq q \leq \frac{\omega}{v_{\rm F}}~,
\ee
and zero elsewhere.

\subsection{Intra-band scattering and the collinear scattering singularity}

We clearly see from Eq.~(\ref{eq:final_angular_integral_MDF_intraband}) that the denominator of $A_{++}(q,k,\omega)$ vanishes like $\sqrt{q^2 - \omega^2/v^2_{\rm F}}$ for $\omega \to \pm v_{\rm F} q$.
On the contrary, the zero in the denominator of the inter-band angular factor $A_{+-}(q,k,\omega)$ is cancelled by the coherence factor ${\cal F}^\star_{+-}(k,q,\omega)$, which vanishes as $q^2 - \omega^2/v^2_{\rm F}$ for $\omega \to \pm v_{\rm F} q$. 

Now, the imaginary part of the non-interacting density-density response function, $\Im m [\chi^{(0)}(q,\omega,T)]$, in Eq.~(\ref{eq:quasielectron-quasihole-lifetime}) diverges~\cite{ramezanali_jphysa_2009} like $1/\sqrt{q^2 - \omega^2/v^2_{\rm F}}$. The combination of these two facts produces an overall factor $1/(q^2  - \omega^2/v^2_{\rm F})$ in Eq.~(\ref{eq:quasielectron-quasihole-lifetime}). Because of this factor,  the standard static screening approximation,\cite{giuliani_prb_1982,Giuliani_and_Vignale} which consists in replacing $\varepsilon(q,\omega,T)$ by $\varepsilon(q,0,T)$ in Eq.~(\ref{eq:quasielectron-quasihole-lifetime}), is seen to  fail miserably in doped graphene, yielding  a logarithmically-divergent intra-band scattering rate~\cite{kashuba_prb_2008,fritz_prb_2008}.   The divergence arises from the regions of phase space in which $\omega =  \pm v_{\rm F} q$. This condition characterizes scattering events in which all the involved electronic momenta are parallel to each other. The ``collinear scattering" singularity has been known for a long time in systems with linear-in-momentum energy bands (see, for example, Ref.~\onlinecite{sachdev_prb_1998}) and has been extensively discussed in the recent graphene-related literature (see, for example, Ref.~\onlinecite{tomadin_prb_2013} and references therein to earlier work). This divergence can be handled in a variety of ways: one can, for example, introduce a cut-off in the integration over $q$ or use dynamical screening, as the G$_0$W-RPA theory we have adopted since the very beginning seems to suggest. Dynamical RPA screening, indeed, naturally cures the collinear scattering singularity because $\varepsilon(q,\omega,T)$ in Eq.~(\ref{eq:quasielectron-quasihole-lifetime}) diverges precisely as $1 / \sqrt{q^2  - \omega^2/v^2_{\rm F}}$ upon approaching $ q =  |\omega|/v_{\rm F}$.

\subsection{Asymptotic behavior of the decay rate for weakly-excited states}

With this body of knowledge at our disposal, we can now evaluate Eq.~(\ref{eq:quasielectron-quasihole-lifetime}) analytically. 
For the sake of definiteness, we consider an $n$-doped graphene sheet with electron density $n$, Fermi energy 
$\varepsilon_{\rm F} = \hbar v_{\rm F} k_{\rm F}$, and Fermi momentum $k_{\rm F} = \sqrt{4\pi n/N_{\rm f}}$. Here $N_{\rm f} =4$ is the number of fermion flavors.

Let us start by considering the first term in Eq.~(\ref{eq:quasielectron-quasihole-lifetime}), i.e.~the quasiparticle decay rate. 
We consider a weakly-excited state composed of a quasiparticle with $k \gtrsim k_{\rm F}$ and $k_{\rm B} T \ll \varepsilon_{\rm F}$. 
The thermal factor
\be
\frac{1- n_{\rm F}(\xi_{{\bm k},+} - \omega)}{1 - \exp(-\beta\omega)}
\ee
imposes some natural bounds on the integration domain with respect to the energy transfer $\omega$. For $\omega < 0$, the natural lower bound of integration is $k_{\rm B}T$. For $\omega>0$ the upper bound of integration is of the order of $\xi_{{\bm k},+} = v_{\rm F}(k-k_{\rm F}) \ll \varepsilon_{\rm F}$. For $k \to k_{\rm F}$, moreover, we can approximate Eqs.~(\ref{eq:final_angular_integral_MDF_intraband})-(\ref{eq:final-inequality-intraband}) as
\ber\label{eq:App}
A_{++}(k,q,\omega) &\simeq& \frac{2c (1- q^2/4k^2_{\rm F})}{v_{\rm F}\sqrt{(1 - q^2/4k^2_{\rm F})(q^2 -\omega^2/v^2_{\rm F})}}\nonumber\\
&=&\frac{2c\sqrt{1 - q^2/4k^2_{\rm F}}}{v_{\rm F}\sqrt{q^2 -\omega^2/v^2_{\rm F}}}~.
\eer
From now on, we set $c=1$ in Eq.~(\ref{eq:App}) since $c \neq 1$ on a set of zero measure with respect to the 2D integral in Eq.~(\ref{eq:quasielectron-quasihole-lifetime}). 

In the same limits, 
the inter-band contribution to the quasiparticle decay rate vanishes since, on the one hand, the thermal factor $1- n_{\rm F}(\xi_{{\bm k},+} - \omega)$ imposes $\omega< \xi_{{\bm k}, +}$ for $T\to 0$, while, on the other hand, $A_{+-}(q,k,\omega)$ is non zero if and only if $\omega \geq v_{\rm F} k$ (at any temperature---see Sect.~\ref{sect:inter-band}). Finally, we emphasize that {\it only} the spectral density of intra-band electron-hole pairs contributes to $1/\tau^{({\rm e})}_{{\bm k}, +}$, since $A_{++}(k,q,\omega) \neq 0$ if and only if $q> |\omega|/v_{\rm F}$.

For small values of the energy transfer $\omega$, $|\omega|/v_{\rm F} \leq q \leq 2k_{\rm F} - \omega/v_{\rm F}$, and $k_{\rm B} T\ll \varepsilon_{\rm F}$ the imaginary part of the non-interacting density-density response function 
can be approximated as following:
\ber\label{eq:approximateimchizero}
\Im m[\chi^{(0)}(q,\omega,T)] &\simeq& -N(0)\frac{\omega}{v_{\rm F}q}\sqrt{1 - \frac{q^2}{4k^2_{\rm F}}}\nonumber\\
&\times&\frac{q}{\sqrt{q^2 - \omega^2/v^2_{\rm F}}}~,
\eer
where
\be\label{eq:DOS}
N(0) = \frac{N_{\rm f}k_{\rm F}}{2\pi v_{\rm F}}
\ee
is the density-of-states at the Fermi energy. Eq.~(\ref{eq:approximateimchizero}) is a contribution of purely intra-band origin. Note that: i) $\Im m[\chi^{(0)}(q,\omega,T)]$ is proportional (and not inversely proportional, as in the ordinary 2DEG~\cite{Giuliani_and_Vignale}) to the factor $\sqrt{1 - q^2/4k^2_{\rm F}}$: this fact beautifully reflects the impossibility of MDFs to be backscattered; ii) we have retained the frequency dependence of the factor on the second line of the previous equation: this is crucial to regularize the collinear scattering singularity for $q \to |\omega|/v_{\rm F}$ in Eq.~(\ref{eq:App}).

In the same range of values of $\omega$, $q$, and $T$ we have
\be
\Re e[\chi^{(0)}(q,\omega,T)] = - N(0)~.
\ee
We therefore conclude that, in the relevant range of values of $\omega$, $q$, and $T$, the RPA dielectric function can be well approximated by
\ber\label{eq:screening}
\varepsilon(q,\omega,T) &\simeq& 1+\frac{2\pi e^2 N(0)}{\epsilon q} - i \frac{2\pi e^2 N(0)}{\epsilon q}\nonumber\\
&\times&\frac{\omega}{v_{\rm F}q}\sqrt{1 - \frac{q^2}{4k^2_{\rm F}}}\frac{q}{\sqrt{q^2 - \omega^2/v^2_{\rm F}}}~.
\eer
It is useful at this stage to introduce the Thomas-Fermi screening wave vector:
\be\label{eq:TF}
q_{\rm TF} \equiv \frac{2\pi e^2 N(0)}{\epsilon} = N_{\rm f} \alpha_{\rm ee} k_{\rm F}~.
\ee

We therefore find
\begin{widetext}
\ber
\frac{1}{\tau^{({\rm e})}_{{\bm k},+}} &\simeq& \frac{4N(0)}{(2\pi)^2v^2_{\rm F}}\int_{-\infty}^{+\infty}d\omega~\omega\frac{1- n_{\rm F}(\xi_{{\bm k},+} - \omega)}{1 - \exp(-\beta\omega)}\nonumber \\ 
&\times&\int_{|\omega|/v_{\rm F}}^{2k_{\rm F}-\omega/v_{\rm F}}dq~q\frac{v^2_q}{\displaystyle\left(1 +\frac{q_{\rm TF}}{q}\right)^2 + \frac{q^2_{\rm TF}}{q^2}\frac{\omega^2}{v^2_{\rm F}}\frac{1 - q^2/4k^2_{\rm F}}{q^2 - \omega^2/v^2_{\rm F}}}\frac{1 - q^2/4k^2_{\rm F}}{q^2 -\omega^2/v^2_{\rm F}}\nonumber\\
&=&4N(0)\alpha^2_{\rm ee}\int_{-\infty}^{+\infty}d\omega~\omega\frac{1- n_{\rm F}(\xi_{{\bm k},+} - \omega)}{1 - \exp(-\beta\omega)}\nonumber\\
&\times&\int_{|\omega|/v_{\rm F}}^{2k_{\rm F} - \omega/v_{\rm F}}dq~\frac{1}{q}\frac{1}{\displaystyle\left(1 +\frac{q_{\rm TF}}{q}\right)^2 + \frac{q^2_{\rm TF}}{q^2}\frac{\omega^2}{v^2_{\rm F}}\frac{1 - q^2/4k^2_{\rm F}}{q^2 - \omega^2/v^2_{\rm F}}}\frac{1 - q^2/4k^2_{\rm F}}{q^2 -\omega^2/v^2_{\rm F}}~.
\eer
\end{widetext}
The integral over $q$ in the previous equation is easily seen to diverge logarithmically for $\omega \to 0$. Indeed, we can estimate the integral over $q$ as follows: 
\begin{widetext}
\ber\label{eq:integraloverq}
&&\int_{|\omega|/v_{\rm F}}^{2k_{\rm F} - \omega/v_{\rm F}}dq~\frac{1}{q}\frac{1}{\displaystyle\left(1 +\frac{q_{\rm TF}}{q}\right)^2 + \frac{q^2_{\rm TF}}{q^2}\frac{\omega^2}{v^2_{\rm F}}\frac{1 - q^2/4k^2_{\rm F}}{q^2 - \omega^2/v^2_{\rm F}}}\frac{1 - q^2/4k^2_{\rm F}}{q^2 -\omega^2/v^2_{\rm F}} \simeq \frac{1}{q^2_{\rm TF}}\ln{\left(\frac{\Lambda}{|\omega|}\right)}~,
\eer
\end{widetext}
where $\Lambda$ is an arbitrary ultraviolet cut-off whose value does not affect the results to {\it leading} order in the low-energy and low-temperature limits. 

To obtain Eq.~(\ref{eq:integraloverq}) we have neglected the first term in the denominator, which is much smaller than the second term since the latter diverges as $(q^2 - \omega^2/v^2_{\rm F})^{-1}$ when $q$ approaches the lower bound of integration. In other words, ``dynamical overscreening'', 
which occurs near the light cone $\omega = \pm v_{\rm F} q$ of a MDF system, completely dominates over the conventional static screening $(1+q_{\rm TF}/q)^2$. From Eq.~(\ref{eq:integraloverq}) it is clear that the logarithmic divergence for $|\omega| \to 0$ originates from the region of small momenta. In the ordinary 2DEG, a similar divergence~\cite{Giuliani_and_Vignale} picks a finite contribution also from the region $q \sim 2k_{\rm F}$. Chirality of MDFs strongly suppresses this contribution in the case of a doped graphene sheet.  

In summary, we find
\be\label{eq:summary}
\frac{1}{\tau^{({\rm e})}_{{\bm k},+}} \to \frac{4N(0)}{N^2_{\rm f} k^2_{\rm F}}\int_{-\infty}^{+\infty}d\omega~\omega\ln{\left(\frac{\Lambda}{|\omega|}\right)}\frac{n_{\rm F}(\omega- \xi_{{\bm k},+})}{1 - \exp(-\beta\omega)}~,
\ee
where we have used that $n_{\rm F}(x) + n_{\rm F}(-x) = 1$. Note that the final result (\ref{eq:summary}) does not depend on the fine-structure constant $\alpha_{\rm ee}$. As emphasized in the Introduction, this feature arises from the fact that the dominant contribution to the quasiparticle decay rate arises from scattering processes with small momentum transfer $q$. For these processes the Coulomb interaction between two quasiparticles is strongly screened by the electronic medium, leading to an effective interaction that depends only on the non-interacting density of states $N(0)$.

The above Eq.~(\ref{eq:summary}) is valid regardless of the relative magnitude of temperature and quasiparticle energy $\xi_{{\bm k}}$, provided they are both much smaller than the Fermi energy.  We now specialize to the case in which one of these two quantities is much larger than the other.   Following Ref.~\onlinecite{Giuliani_and_Vignale}, we first consider the zero-temperature limit in which $\beta \xi_{{\bm k}, +}\gg 1$. In this case the main contribution to the previous integral comes from the region $\omega \sim \xi_{{\bm k}, +}$. Since the logarithm is a slowly-varying function of its argument we find
\be\label{eq:final-result-zero-temperature}
\frac{1}{\tau^{({\rm e})}_{{\bm k},+}}\simeq\frac{4N(0)}{N^2_{\rm f} k^2_{\rm F}}\ln{\left(\frac{\Lambda}{\xi_{{\bm k}, +}}\right)}\int_{-\infty}^{+\infty}d\omega~\omega\frac{n_{\rm F}(\omega - \xi_{{\bm k},+})}{1 - \exp(-\beta\omega)}~.
\ee
We now use the ``beautiful integral" (Ref.~\onlinecite{Giuliani_and_Vignale}, p. 497)
\be\label{eq:famous-integral}
\int_{-\infty}^{+\infty}dy \frac{x-y}{(1+ e^{-y})(1 - e^{y-x})} = \frac{x^2 +\pi^2}{2(1+ e^{-x})}~,
\ee
with $x = \beta\xi_{{\bm k}, +}$ and $y = \beta(\xi_{{\bm k}, +} - \omega)$. We find
\be\label{eq:famous-integral-specialized}
\int_{-\infty}^{+\infty}d\omega~\omega\frac{n_{\rm F}(\omega -\xi_{{\bm k},+})}{1 - \exp(-\beta\omega)} = \frac{1}{\beta^2}\frac{ \beta^2\xi^2_{{\bm k}, +}+\pi^2}{2(1+ e^{- \beta\xi_{{\bm k}, +}})}~.
\ee
In the regime $\beta \xi_{{\bm k}, +} \gg 1$ we therefore have
\be\label{eq:final-result-zero-temperature}
\frac{1}{\tau^{({\rm e})}_{{\bm k},+}} \simeq \frac{\varepsilon_{\rm F}}{\hbar}\frac{1}{\pi N_{\rm f}}\left(\frac{\xi_{{\bm k}, +}}{\varepsilon_{\rm F}}\right)^2\ln{\left(\frac{\Lambda}{\xi_{{\bm k}, +}}\right)}~,
\ee
where we have restored $\hbar$. A careful inspection of Eq.~(\ref{eq:quasielectron-quasihole-lifetime}) shows that, in the limit $\beta |\xi_{{\bm k}, +}| \gg 1$, the decay rate of a plane-wave state is given by
is
\be\label{eq:PW_lifetime_zeroT}
\frac{1}{\tau_{{\bm k},+}} \simeq \frac{\varepsilon_{\rm F}}{\hbar}\frac{1}{\pi N_{\rm f}}\left(\frac{\xi_{{\bm k}, +}}{\varepsilon_{\rm F}}\right)^2\ln{\left(\frac{\Lambda}{|\xi_{{\bm k}, +}|}\right)}~,
\ee
i.e.~it is equal to that in the right-hand side of Eq.~(\ref{eq:final-result-zero-temperature}), provided that we replace $\xi_{{\bm k}, +} \to |\xi_{{\bm k}, +}|$ in this equation.
To leading order in the zero-temperature limit the functional dependence on $\xi_{{\bm k}, +}$, i.e.~$-\xi^2_{{\bm k}, +}\ln(|\xi_{{\bm k},+}|)$, coincides with that of an ordinary 2DEG~\cite{giuliani_prb_1982}.

We now turn to analyze the other relevant regime, i.e.~$\beta|\xi_{{\bm k}, +}|\ll 1$. In this case the main contribution to the integral comes from a region of the order of $k_{\rm B}T$ centered around the origin. Once again, the logarithm can be taken out of the integral giving a factor $\ln{(\Lambda/k_{\rm B} T)}$. 

In the regime $\beta|\xi_{{\bm k}, +}|\ll 1$ we therefore find
\ber\label{eq:final-result-finite-temperature}
\frac{1}{\tau^{({\rm e})}_{{\bm k},+}} \simeq  \frac{\varepsilon_{\rm F}}{\hbar} \frac{\pi}{2N_{\rm f}} \left(\frac{k_{\rm B} T}{\varepsilon_{\rm F}}\right)^2\ln{\left(\frac{\Lambda}{k_{\rm B} T}\right)}~,
\eer
where we have used an expansion of Eq.~(\ref{eq:famous-integral-specialized}) for $\beta|\xi_{{\bm k}, +}| \to 0$ and we have restored $\hbar$. 
Inspecting Eq.~(\ref{eq:quasielectron-quasihole-lifetime}) for $\xi_{{\bm k}, +} =0$ we conclude that 
\be\label{eq:PW_lifetime_finiteT}
\lim_{\xi_{{\bm k}, +} \to 0} \frac{1}{\tau_{{\bm k},+}}  = \frac{2}{\tau^{({\rm e})}_{{\bm k},+}} \simeq \frac{\varepsilon_{\rm F}}{\hbar} \frac{\pi}{N_{\rm f}} \left(\frac{k_{\rm B} T}{\varepsilon_{\rm F}}\right)^2
\ln{\left(\frac{\Lambda}{k_{\rm B} T}\right)}~.
\ee
Once again, the functional dependence on temperature of Eq.~(\ref{eq:final-result-finite-temperature}), i.e.~$-T^2\ln(T)$, coincides with that obtained by Giuliani and Quinn~\cite{giuliani_prb_1982} for an ordinary 2DEG.

\section{Summary and conclusions}

In summary, we have presented a pedagogical derivation of the quasiparticle lifetime in a doped graphene sheet. Three main differences with respect to the classic Giuliani-Quinn calculation~\cite{giuliani_prb_1982} for an ordinary two-dimensional electron gas have been identified: i) a simple Fermi golden rule approach with statically screened Coulomb interactions is not viable in graphene as it yields logarithmically-divergent intra-band scattering rates due to the collinear scattering singularity; ii) the leading-order contribution to the quasiparticle decay rate in the low-energy and low-temperature limits is completely controlled by scattering events with small momentum transfer: 
$2k_{\rm F}$ contributions are suppressed by the chiral nature of massless Dirac carriers in graphene; iii) because of ii), the leading order contribution to the quasiparticle decay rate is completely independent on the strength on the background dielectric constant $\epsilon$: the result is therefore {\it universal} in that it does not depend on the substrate on which graphene is placed.

Finally, we emphasize how the recently developed ability to align the crystals of two graphene sheets~\cite{novoselov_private_commun} paves the way for two-dimensional-to-two-dimensional tunneling experiments~\cite{murphy_prb_1995} 
in which inter-layer tunneling does not spoil momentum conservation. These experiments may allow a direct measurement of the temperature and doping dependence of the 
quasiparticle lifetime in high-quality graphene sheets.

\acknowledgements
We gratefully acknowledge Leonid Levitov, Kostya Novoselov, Alessandro Principi, Justin Song, and Andrea Tomadin for useful discussions. This Article is dedicated to the memory of our friend Gabriele F. Giuliani,  physicist, soccer player, car racer and provocateur, who passed away on November 22, 2012, after a heroic battle with a very aggressive form of cancer that lasted 12 years.

\end{document}